\documentclass{aa}
\usepackage{amsmath}
\usepackage{graphicx}
\usepackage{natbib}
\bibpunct{(}{)}{;}{a}{}{,}
\usepackage{txfonts}
\newcommand{\kms}{\,km\,s$^{-1}$}
\newcommand{\myr}{\,$M_{\sun}\,{\rm yr}^{-1}$}
\newcommand{\ro}{\,$R_{\sun}$}
\newcommand{\mo}{\,$M_{\sun}$}
\newcommand{\lo}{\,$L_{\sun}$}
\begin{document}
%
\title{Wind mass transfer in S-type symbiotic binaries}
\subtitle{I. Focusing by the wind compression model}
\author{A. Skopal \and Z. Carikov\'{a}}
\institute{Astronomical Institute, Slovak Academy of Sciences,
           059~60 Tatransk\'{a} Lomnica, Slovakia} 
		
\date{Received / Accepted}
\abstract
{
Luminosities of hot components in symbiotic binaries require 
accretion rates that are higher than those that can be achieved 
via a standard Bondi-Hoyle accretion. This implies that the wind 
mass transfer in symbiotic binaries has to be more efficient. 
}
{
We suggest that the accretion rate onto the white dwarfs (WDs) 
in S-type symbiotic binaries can be enhanced sufficiently by 
focusing the wind from their slowly rotating normal 
giants towards the binary orbital plane. 
}
{
We applied the wind compression model to the stellar wind of 
slowly rotating red giants in S-type symbiotic binaries. 
}
{
Our analysis reveals that for typical terminal velocities of 
the giant wind, 20 to 50\kms, and measured rotational velocities 
between 6 and 10\kms, the densities of the compressed wind at 
a typical distance of the accretor from its donor correspond 
to the mass-loss rate, which can be a factor of $\sim$10 higher 
than for the spherically symmetric wind. This allows the WD to 
accrete at rates of $10^{-8} - 10^{-7}$\myr, and thus to power 
its luminosity. 
}
{
We show that the high wind-mass-transfer efficiency in S-type 
symbiotic stars can be caused by compression of the wind from 
their slowly rotating normal giants, whereas in D-type symbiotic 
stars, the high mass transfer ratio can be achieved via the 
gravitational focusing, which has recently been suggested for 
very slow winds in Mira-type binaries. 
}

\maketitle
\section{Introduction}

Symbiotic stars are long-period interacting binary systems 
that include a cool giant as the donor star and a compact 
star, in most cases a white dwarf (WD), as the accretor. 
On the basis of their infrared properties, we distinguish 
between S-type (Stellar) and D-type (Dusty) symbiotic stars 
\citep[][]{wa75}. The former is represented by a stellar 
type of the IR continuum from a normal giant, whereas the 
latter contains additional strong emission from the dust 
produced by a Mira-type variable. 
For D-type systems, the orbital periods are, in general, 
unknown, but considered to be in the range of a few 
times 10 to 100 years. The cases of V1016~Cyg 
\citep[$P_{\rm orb}\sim 15~{\rm years}$,][]{parimucha+02} 
and Mira AB with the projected separation of components of 
$\sim 65$\,AU \citep[][]{matt+kar06} represent 
the well-measured examples. 
Most of S-type systems has orbital periods between 200 and 
1000 days \citep[e.g.][]{belcz+00}, which correspond to the 
binary extension of a few AU for their typical total mass 
of $\approx 2$\mo\ \citep[e.g.][]{schild+96}. 
Based on a strong correlation between the spectral type of 
the cool giant and the orbital period, \cite{ms99} suggested 
that symbiotic stars are well detached binary systems. 
Thus the symbiotic activity on the WD companion is 
triggered by the wind mass transfer. 

The WD therefore accretes a fraction of the giant wind, which 
leads to its heating up to $\ga 10^5$\,K and increasing its 
luminosity to $\sim 10^{2}-10^{4}$\lo. This hot WD then 
ionizes a portion of the wind from the giant giving rise to 
the nebular radiation \citep[e.g.][]{stb}. 
Without any sudden optical brightenings, this configuration 
is called the quiescent phase of a symbiotic star. 
In most cases, the observed large energy output is believed 
to be caused by stable nuclear hydrogen burning on the WD 
surface \citep[e.g.][]{tutukov+76}, which requires accretion 
onto a low mass WD at rates of $10^{-8} - 10^{-7}$\myr\ 
\citep[see Fig.~2 of][]{shen+07}. 
However, such high accretion rates cannot be achieved by 
a standard Bondi-Hoyle wind accretion, because of its low 
efficiency (= the mass transfer ratio) of a few percent 
\citep[][]{b+h44,livwar84} and the mass-loss rates from red 
giants in S-type systems of $\approx 10^{-7}$\myr\ 
\citep[e.g.][]{m+91,mio02,sk05,walder+08}. 
Similar results were obtained by even more sophisticated 
three-dimensional simulations of wind accretion in well 
separated binaries. Depending on the binary configuration, 
the mass transfer ratio (= mass accretion rate/mass loss rate) 
was found to be in the range of 0.6--10\% 
\citep[e.g.][]{tbj96,walder97,dumm+00,nagae+04,walder+08}. 

The long-standing problem of the large energetic output from 
hot components in symbiotic binaries and its deficient fuelling 
by the giant in the canonical Bondi-Hoyle picture was first 
pointed out by \cite{ken+gall83}. 
A promising solution of the problem has recently been suggested 
for Mira-type interacting binaries, whose slow and dense winds 
from evolved AGB giants can be gravitationally focused on 
the binary orbital plane \citep[][]{moh+pod07,moh+pod12,borro+09}. 
This mass transfer mode, where the wind is filling the Roche 
lobe instead of the star itself, is called wind Roche-lobe 
overflow \citep[WRLOF,][]{moh+pod07}. In the model, material 
is gravitationally confined to the Roche lobe and falls 
into the potential of the companion through the $L_1$ point, 
similarly to standard RLOF. In general, WRLOF can occur in systems 
where the wind acceleration zone lies close to, or is a significant 
fraction of, the Roche lobe radius. Corresponding mass transfer
efficiencies are at least an order of magnitude more than 
the analogous Bondi-Hoyle values \citep[][]{moh+pod12}. 

In this contribution we suggest that the wind from normal giants 
in S-type symbiotic stars can be focused by their rotation 
according to the wind compression disk (WCD) model elaborated by 
\cite{bjorkcass93}. 
This possibility insists on the fact that normal giants in 
these systems rotate slowly, with the equatorial 
velocity around of 8\kms\ \citep[e.g.][]{zam+07}. In addition, 
a wind-focusing effect was also proven observationally 
\citep[][]{blind+11,boffin+14a,boffin+14b}. 
Accordingly, Sect.~2 outlines the WCD model, and Sect.~3 
presents results of its application. Their discussion and 
conclusion are found in Sects.~4 and 5, respectively. 

\section{The method}

\subsection{Wind compression model}

Rotation of a star with radiation-driven wind leads to its 
compression towards the equatorial regions, where it can create 
a disk-like formation \citep[][]{bjorkcass93}. 
There are two main simplifications in the model: 
  (i) the model neglects the gas pressure gradients at and 
beyond the sonic point, and 
  (ii) it assumes the spherical stars with only the radial 
acting of gravity and radiation forces on wind elements, which 
then move as independent particles ejected with a non-radial 
initial velocity. 
Furthermore, in a star-centred spherical 
coordinate system ($r,\theta,\phi$) with the spin axis 
$\theta = 0$, the WCD model assumes an azimuthal symmetry of 
the wind density. Therefore, it is a function of the radial 
distance $r$ from the star's centre and the polar angle 
$\theta$. According to the mass continuity equation, the 
mass density distribution in the WCD model, 
$\rho^{\rm c}(r,\theta)$, is expressed as 
\begin{equation}
\rho^{\rm c}(r,\theta) = 
          \frac{\dot{M}}{4\pi r^2 v_{\rm r}(r)}
          \left(\frac{d\mu}{d\mu_{0}}\right)^{-1},
\label{eqn:nh}
\end{equation}
where $\dot{M}$ is the mass-loss rate from the star, 
$v_{\rm r}(r)$ is the radial component of the wind velocity 
and the geometrical factor $d\mu/d\mu_{0}$ describes the 
compression of the wind due to rotation of the star 
(see below). 

For the wind velocity profile, we used the $\beta$-law 
as introduced by \cite{lamcass99}, 
\begin{equation}
  v_{\rm r}(r) = v_{\infty}\left(1-\frac{bR_{\star}}{r}
                           \right)^{\beta},
\label{eqn:betalaw}
\end{equation}
where $R_{\star}$ is the radius of the star, $\beta$ 
characterizes acceleration of its wind, and the parameter 
$b$ is given by 
\begin{equation}
  b = 1-\left(\frac{a}{v_{\infty}}\right)^{1/\beta},
\label{eqn:a}
\end{equation}
where $a$ is the initial velocity of the wind at its origin 
$R_{\star}$, and $v_{\infty}$ its terminal velocity. 
In comparison with \cite{bjorkcass93}, we assume that 
$v_{\infty}$ does not depend on the polar angle $\theta$. 

The geometrical factor in Eq.~(\ref{eqn:nh}) is expressed as
\begin{equation}
  \frac{d\mu}{d\mu_{0}} =
        \cos \phi'+\phi' \sin \phi' \cot^{2} \theta_{0} 
\label{eqn:gf}
\end{equation}
\citep[][]{bjorkcass93},
where $\mu = \cos \theta$, $\mu_{0} = \cos \theta_{0}$, and the 
relation between the initial polar angle $\theta_{0}$ of the 
parcel of gas in the streamline (i.e. the polar angle at the 
radius of the star $R_{\star}$) and the actual $\theta$ 
(see Fig.~2 therein) is given by 
\begin{equation}
  \cos \theta = \cos \theta_{0} \cos \phi',
\label{eq:cc}
\end{equation}
where the angle
\begin{equation}
  \phi'(r) = \frac{v_{\rm rot} \sin \theta_{0}}{v_{\infty}}
             \frac{1}{b(1-\beta)}
             \left[\left(1-\frac{bR_{\star}}{r}
             \right)^{1-\beta}-(1-b)^{1-\beta}\right] .
\label{eqn:fi}
\end{equation}
The angle $\phi'$ is called the `displacement angle', which 
determines deflection of a streamline from the radial direction. 
It describes the trajectory of the mass launched from the star's 
surface at $\theta_{0}$ in its own orbital plane. The larger 
the angle $\phi'$, the closer the material gets to the 
equatorial plane of the star. 
For $r \rightarrow \infty$, $\phi'(r)$ has an asymptotic 
value $\phi'_{\rm max}$, which means that the trajectory 
becomes radial at this direction owing to the dominance of 
the radiation forces. Using the $\beta$-law wind as above, 
the asymptotic deflection of the rotating wind particles 
is given by \citep[see][]{lamcass99} 
\begin{equation}
  \phi'_{\rm max} = \frac{v_{\rm rot}\sin \theta_{0}}{v_{\infty}}
              \frac{1}{b(1-\beta)}\left[1-(1-b)^{1-\beta}\right], 
\label{eqn:max}
\end{equation} 
where $v_{\rm rot}$ is the equatorial rotation velocity of 
the star. In our application we consider only models with 
trajectories that do not cross the star's equator 
\citep[the so-called wind compression zone (WCZ) model, see][]
{igncassbjork96}, which means that 
\begin{equation}
  \phi'_{\rm max} < \frac{\pi}{2} .
\label{eqn:wcz}
\end{equation} 
This condition constrains a maximum of $v_{\rm rot}$ for given 
parameters of the stellar wind to form a WCZ. Table~1 gives some 
examples of its value for the wind parameters in Sect.~2.3. 

In our application the rotating star in the model is considered 
to be the red giant (RG) in S-type symbiotic binaries. 
Assuming that the wind from the giant is isothermal at the same 
temperature as the stellar photosphere ($\approx 3000$\,K), 
the speed of sound is a few times \kms. Then, according to 
\cite{bjorkcass93}, the gas pressure gradient force will be 
negligible, because the terminal velocity of the wind and 
the stellar rotation are greater than the isothermal speed 
of sound (parameters in Sects.~2.3 and 2.4). A negligible 
oblateness of the slowly rotating RG \cite[see][]{zcsk14} and 
dropping pressure gradient force result in the purely radial 
radiative drag on wind particles. 
These properties satisfy main simplifications of the WCD model 
(see above), and thus justify its applicability to normal giants 
in S-type symbiotic stars. 

\subsection{Mass-loss ratio}

\begin{figure}
\centering
\begin{center}
\resizebox{\hsize}{!}{\includegraphics[angle=0]{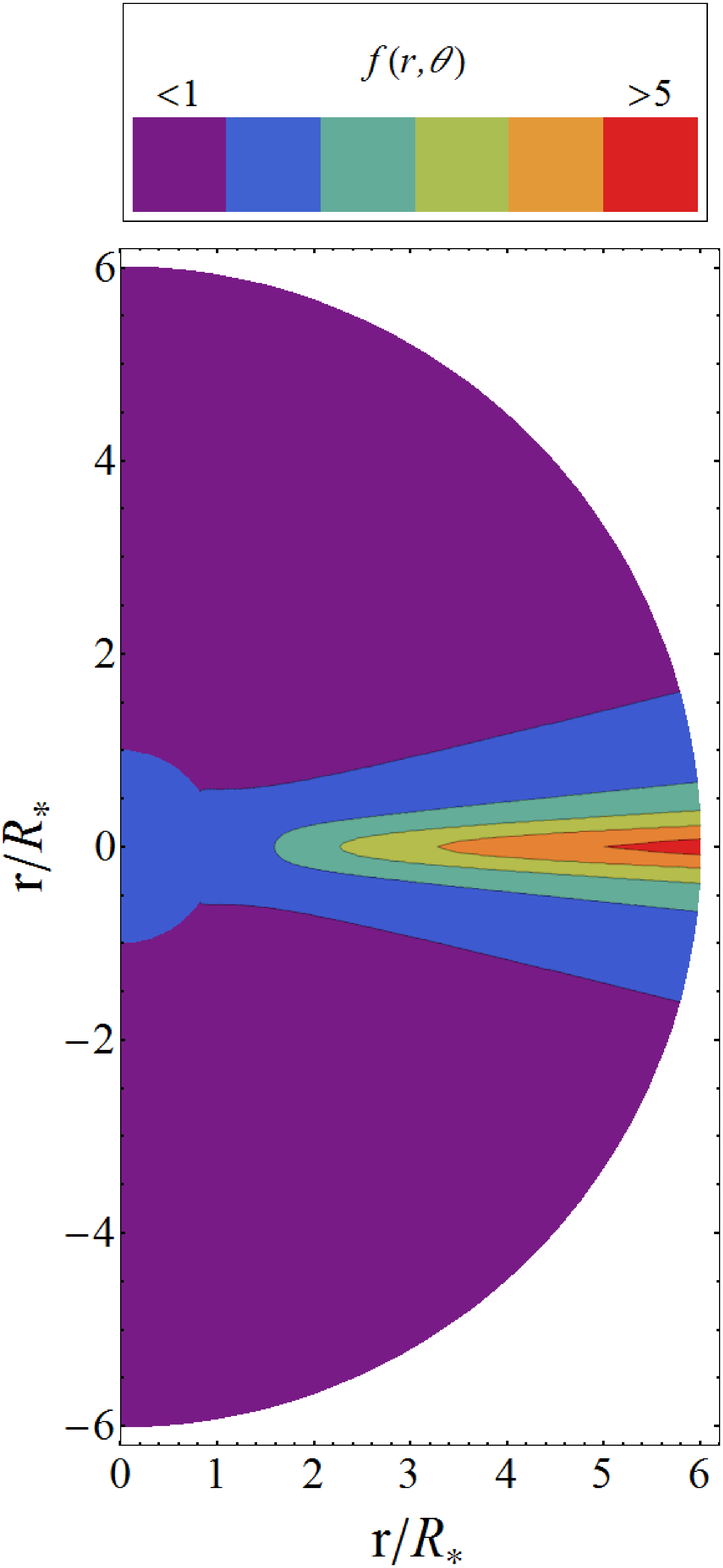}
                      \includegraphics[angle=0]{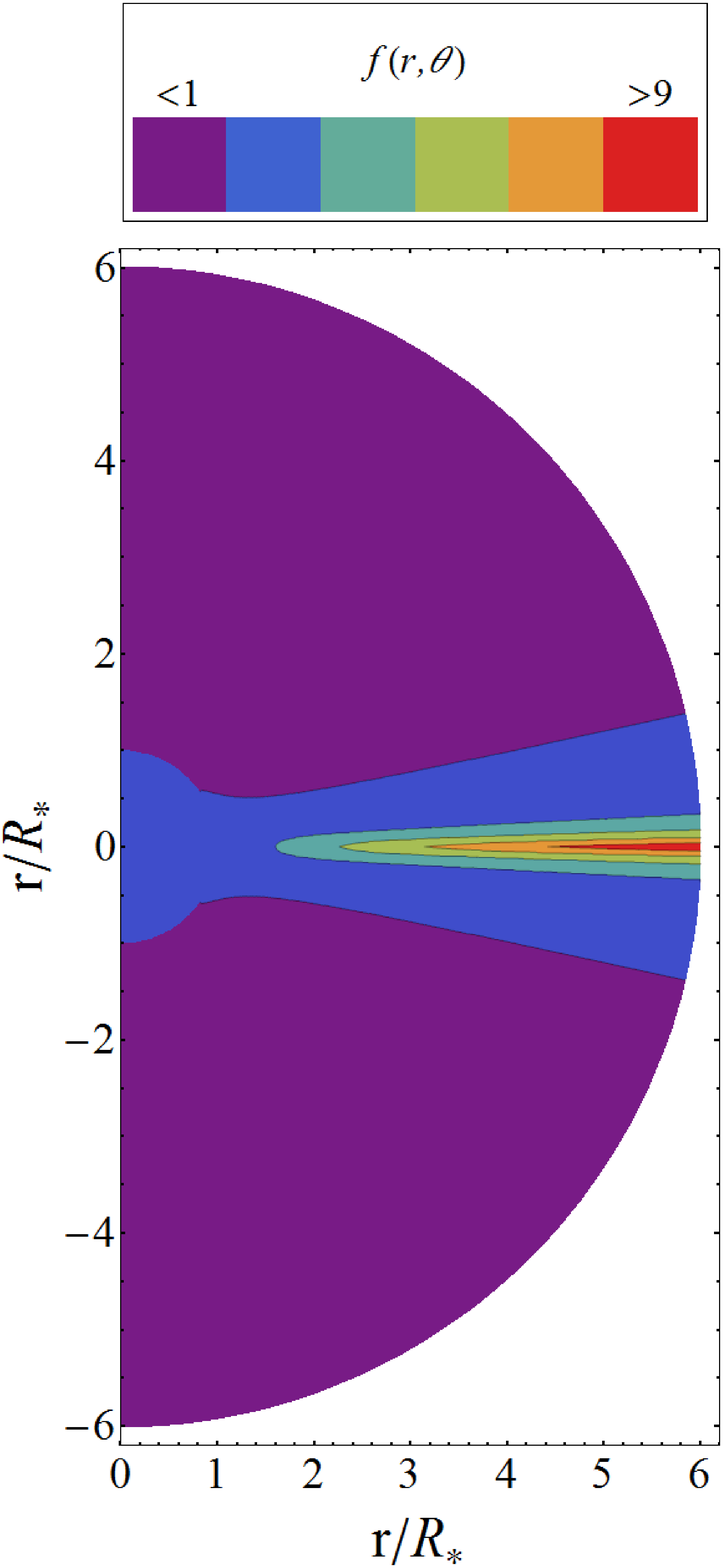}}
\end{center}
\caption[]{
Mass-loss ratio, 
$f(r,\theta) = \dot{M}^{\rm c}(r,\theta)/\dot{M}$, 
calculated for parameters of the model A (left) and 
H (right) in Table~2, depicted in the plane perpendicular 
to the equatorial plane containing the giant at (0,0). 
This demonstrates compression of the wind to the equatorial 
plane with a factor of $ > 5-9$ at $r = 5-6\,R_{\star}$ 
relative to the spherically symmetric density distribution. 
}
\label{fig:f}
\end{figure}

To compare densities of the compressed and spherically symmetric 
wind, we determine mass-loss ratio $f$, which we define as 
the ratio of the mass-loss rate $\dot{M}^{\rm c}(r,\theta)$, 
given by the local density $\rho^{\rm c}(r,\theta)$, 
to the mass-loss rate from the star $\dot{M}$, that is, 
\begin{equation}
 f(r,\theta) = \frac{\dot{M}^{\rm c}(r,\theta)}{\dot{M}} 
             = \frac{\rho^c(r,\theta)}{\rho^{\rm sph}(r)} 
             = \left(\frac{d\mu}{d\mu_{0}}\right)^{-1},
\label{eqn:f}
\end{equation}
where $\rho^{\rm sph}(r)$ is the mass density distribution of 
spherically symmetric wind. Owing to the compression of the wind 
towards the equatorial plane and a constant value of $\dot{M}$, 
the mass-loss ratio $f(r,\theta) > 1$ near the equatorial plane, 
but $< 1$ in directions around the pole. Figure~\ref{fig:f} 
shows two examples of $f(r,\theta)$ for $r < 6\,R_{\star}$ 
and all $\theta$. 

To estimate the accretion from the compressed wind onto the WD, 
we need to determine the mass-loss ratio around the orbital 
plane, which is assumed to coincide with the equatorial plane 
of the rotating star. According to Eq.~(\ref{eqn:f}) it is given 
by the reciprocal value of the geometrical factor (\ref{eqn:gf}), 
which depends on the parameters of the wind and the rotational 
velocity of the star. 
%

\subsection{Parameters of the RG wind}

We assume that the parameters of the wind from the giant 
are characterized by the initial velocity $a = $1\kms, 
terminal velocities $v_{\infty} = 20 - 50$\kms, and the 
acceleration parameter $\beta = 2.5$ \citep[][]{schroder}. 

\subsection{Rotational velocities of RGs}

\citet{soker02} has predicted theoretically that the cool 
components in symbiotic systems are likely to rotate much 
faster than isolated cool giants or those in wide binary 
systems. 
\citet{zam+08} support this suggestion by measuring projected 
rotational velocities, $v_{\rm rot}\sin(i)$, in a number of 
symbiotic stars and isolated giants. They find that in 
S-type symbiotics the K giants rotate two to four times faster 
than the field K giants, and M giants rotate on average 
1.5 times faster than the field M giants. 
Using high-resolution spectroscopic observations and 
the cross-correlation function method, \cite{zam+07} find 
a typical rotational velocity of the K and M giants in S-type 
symbiotic stars to be $4.5 < v_{\rm rot}\sin(i) < 11.7$\kms. 
These results were confirmed by \cite{zamsto12}. Using new data 
for 55 field M0\,III -- M6\,III giants, they indicated a mean 
$v_{\rm rot}\sin(i) = 5.0$\kms\ and median 
$v_{\rm rot}\sin(i) = 4.3$\kms. 
For 33 M0\,III -- M6\,III giants in symbiotic systems, they 
have confirmed a mean $v_{\rm rot}\sin(i) = 7.8$\kms\ and median 
$v_{\rm rot}\sin(i) = 8$\kms. As a result of these studies, we 
assumed the rotational velocities of the cool giants to be in 
the range of 6--9.5\kms. 
%
\begin{figure}
\centering
\begin{center}
\resizebox{\hsize}{!}{\includegraphics[angle=0]{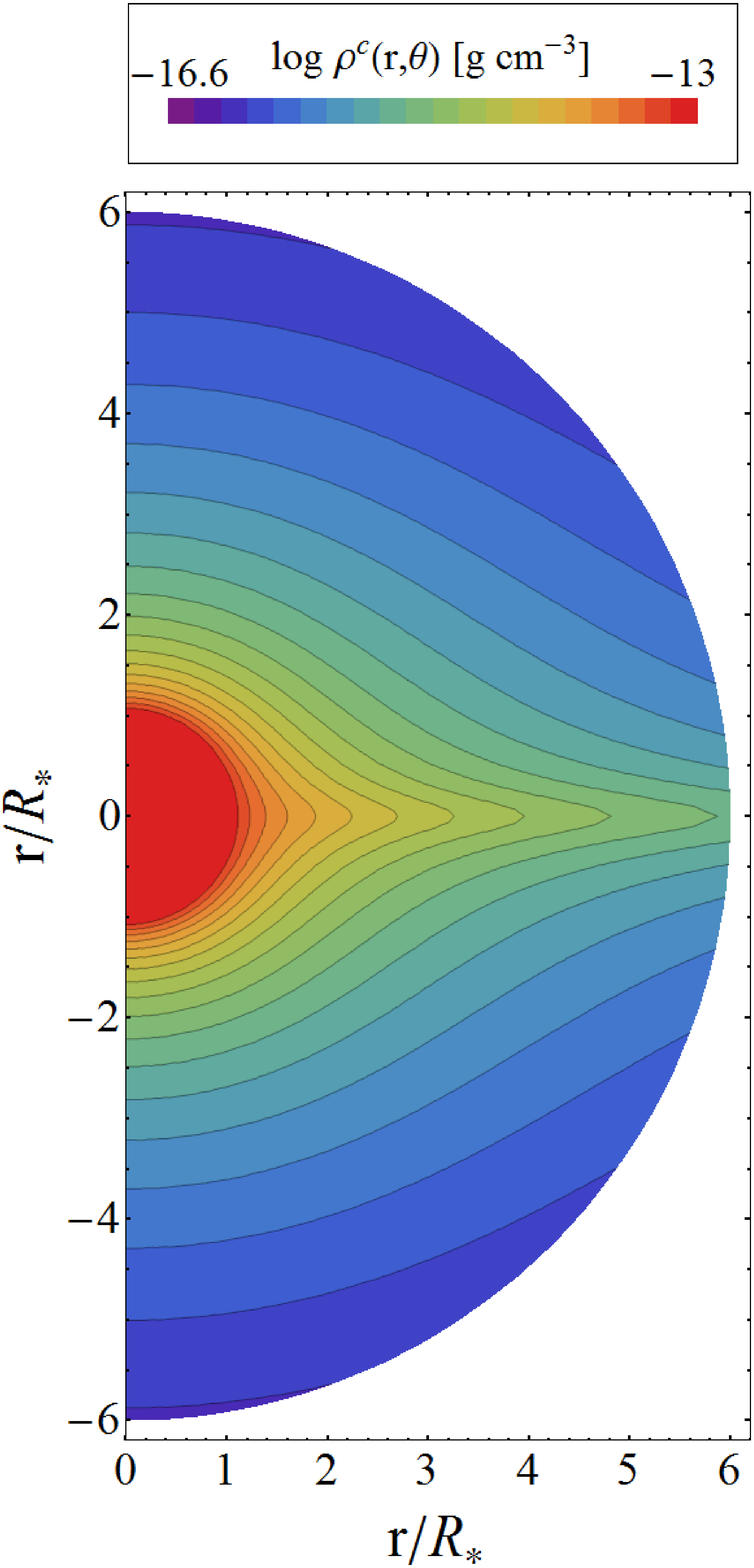}
                      \includegraphics[angle=0]{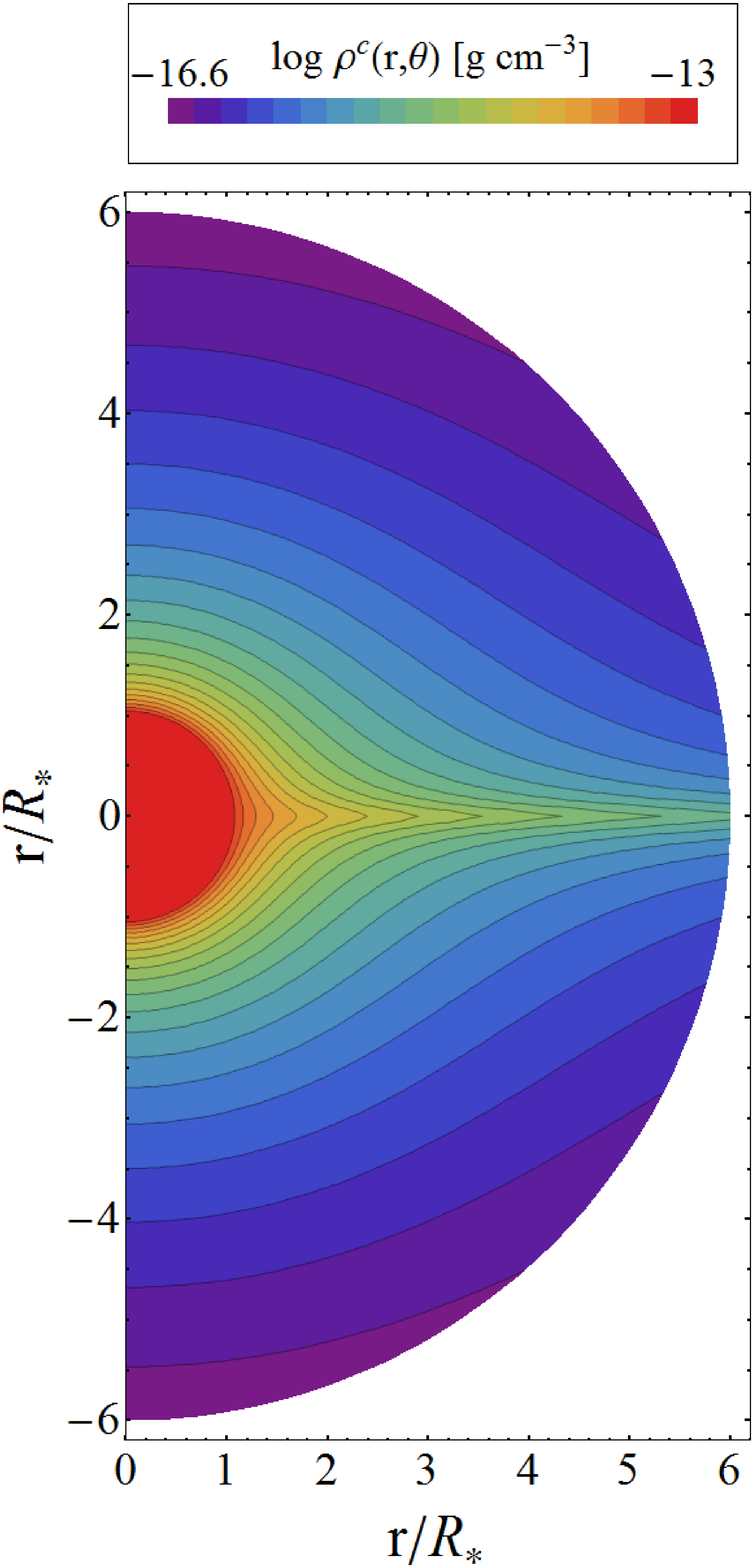}}
\end{center}
\caption[]{
Density distribution in the compressed wind calculated 
according to Eq.~(\ref{eqn:nh}) for $\dot M = 10^{-7}$\myr\ 
and $R_{\star} = 100$\ro. The step of isodensities is 0.17. 
Selection of models and the plane of depiction are 
as in Fig.~\ref{fig:f}. 
          }
\label{fig:den}
\end{figure}
%
%
\begin{figure*}
\centering
\begin{center}
\resizebox{\hsize}{!}{\includegraphics[angle=-90]{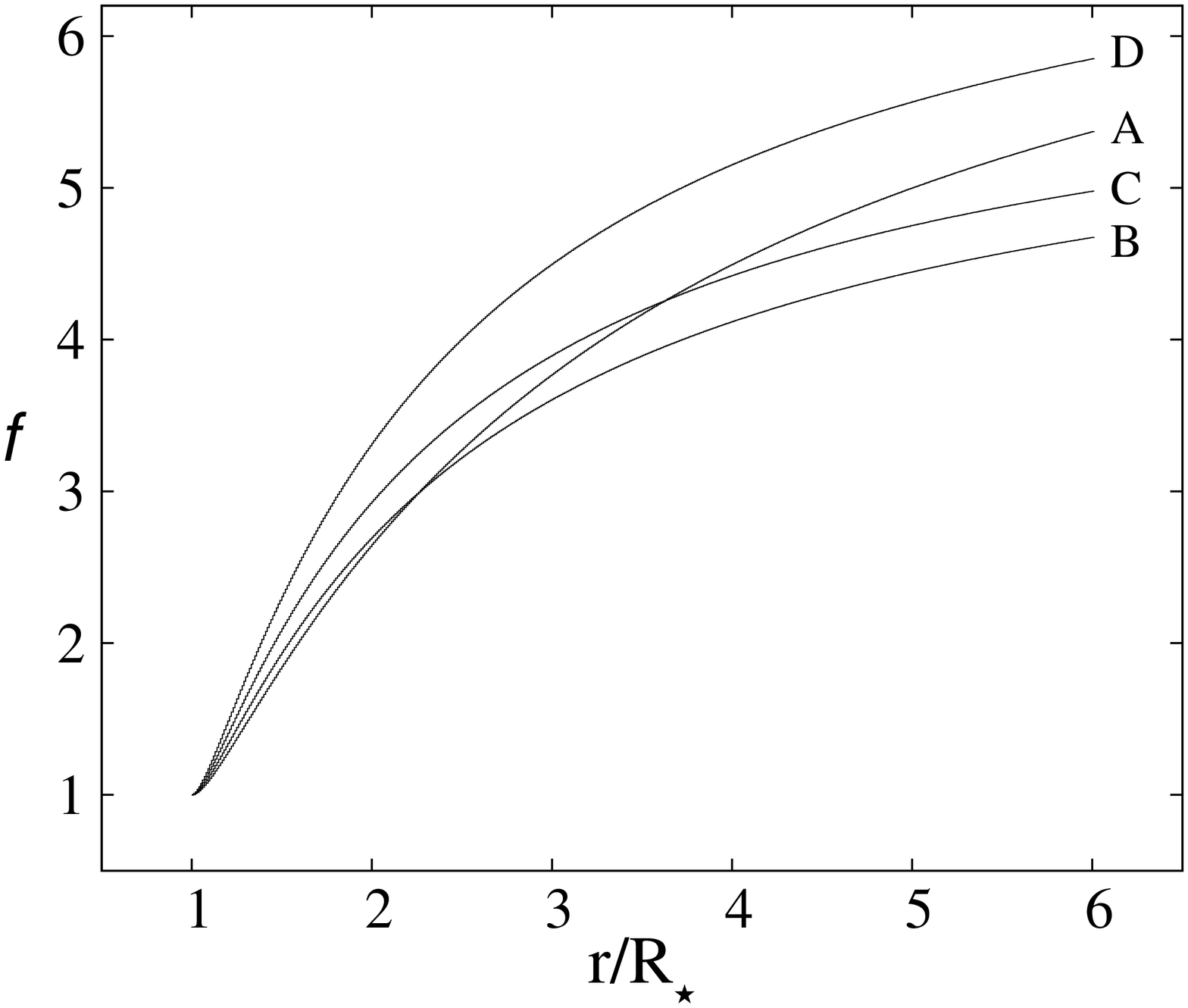}
                      \includegraphics[angle=-90]{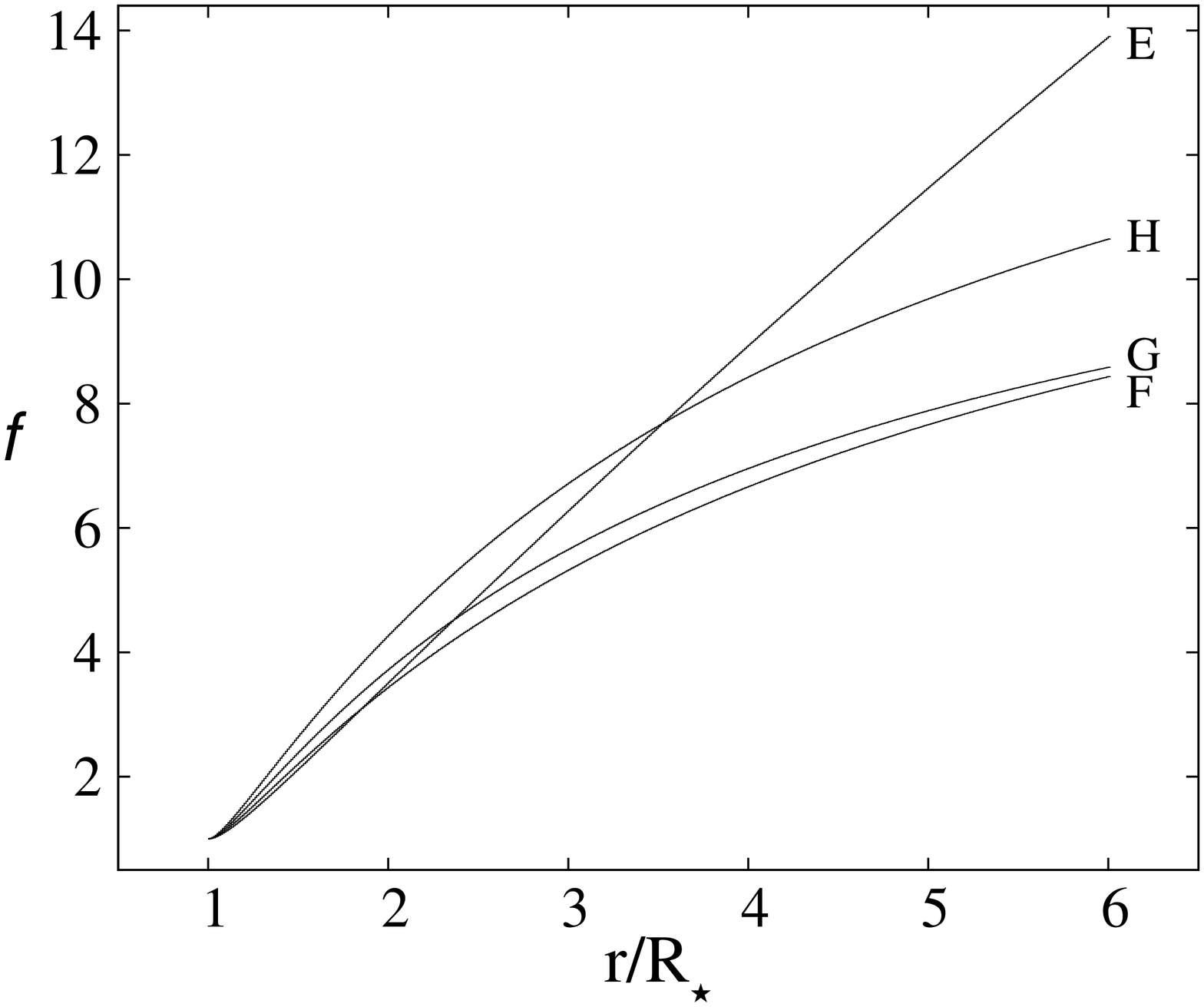}
                      \includegraphics[angle=-90]{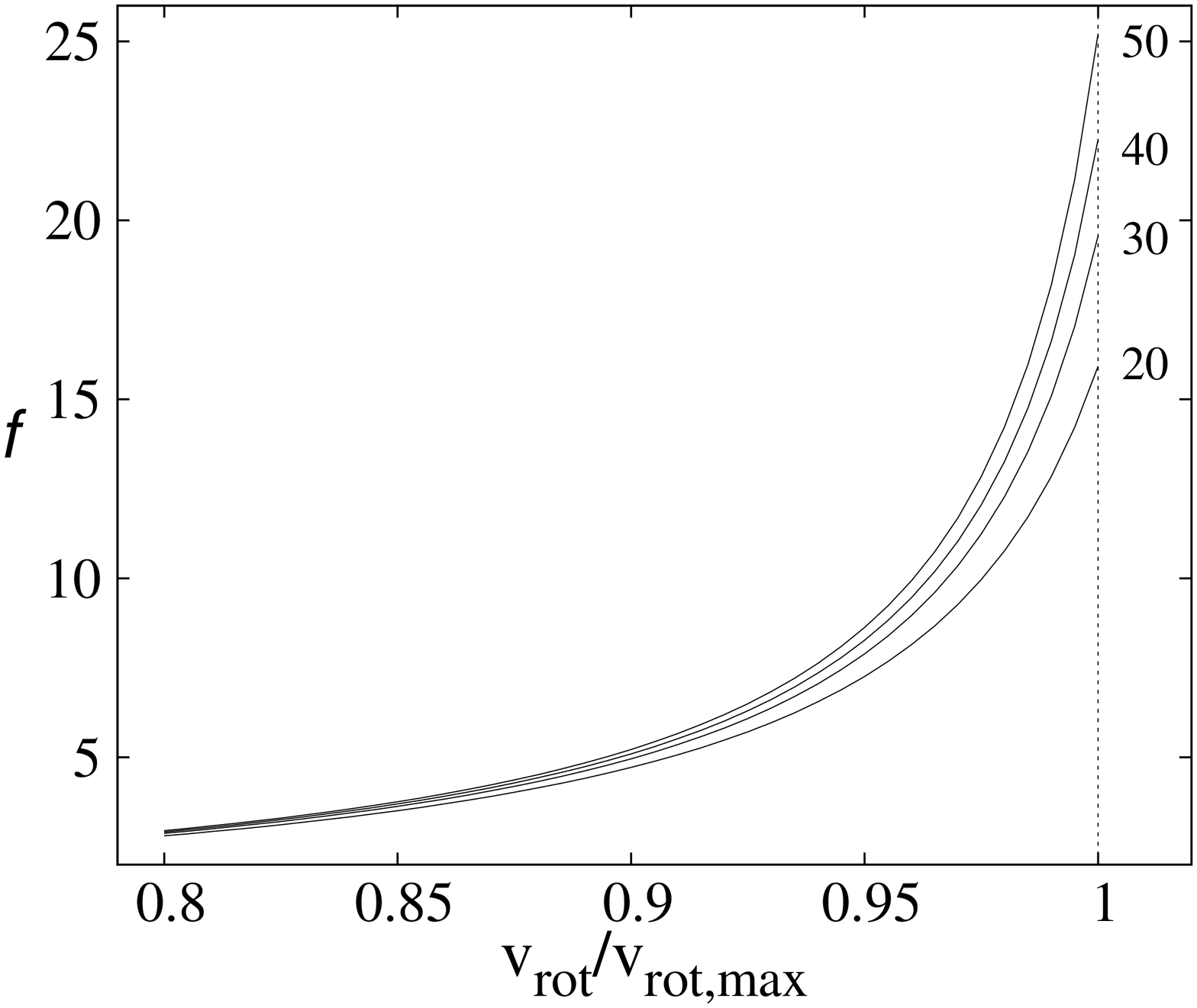}}
\end{center}
\caption[]{
Left and middle panels display the mass-loss ratio $f(r,\pi/2)$ 
at the equatorial plane as a function of the radial distance. 
Parameters of the models are in Table~2. 
The right panel demonstrates an increase of $f(6 R_{\star},\pi/2)$ 
for $v_{\rm rot}/v_{\rm rot,max} \rightarrow 1$ and different 
$v_{\infty}$ (labelled on the right side). 
          }
\label{fig:mm}
\end{figure*}
%
%
\begin{table}
\begin{center}
\caption{
Maximum rotational velocities, $v_{\rm rot,max}$, for non-crossing 
trajectories, calculated according to Eq.~(\ref{eqn:max}) for 
$\phi'_{\rm max} = \pi/2$, $\sin\theta_0 = 1$, $a = 1$\kms, and 
$\beta =2.5$. 
}
%
\begin{tabular}{cc}
\hline
\hline
 $v_{\infty}$ & $v_{\rm rot,max}$ \\
 (\kms)       & (\kms)            \\
\hline
 20 & 6.54  \\
 30 & 7.85  \\
 40 & 8.92  \\
 50 & 9.85  \\
\hline
\end{tabular}
\end{center}
\end{table}
%
\begin{table}
\begin{center}
\caption{
Mass-loss ratio $f$ at $r = 6\,R_{\star}$ and $\theta = \pi/2$ 
calculated for selected values of $v_{\infty}$ and $v_{\rm rot}$ 
(in \kms). Models $f(r,\pi/2)$, denoted here by capital letters, 
are plotted in Fig.~\ref{fig:mm}. 
}
\begin{tabular}{cccc|cccc}
\hline
\hline
 $f$ & $v_{\infty}$ & $v_{\rm rot}$ & model &
 $f$ & $v_{\infty}$ & $v_{\rm rot}$ & model \\
\hline
  5.4 & 20 & 6 & A &   14 & 20 & 6.5 & E \\
  4.7 & 30 & 7 & B &  8.4 & 30 & 7.5 & F \\
  5.0 & 40 & 8 & C &  8.6 & 40 & 8.5 & G \\
  5.9 & 50 & 9 & D &   11 & 50 & 9.5 & H \\
\hline
\end{tabular}
\end{center}
\end{table}

\section{Results}

To demonstrate focusing of the giant's wind towards the equatorial 
plane (assumed to be coincident with the orbital plane) due 
to its rotation, we calculated the mass-loss ratio 
$f(r,\theta)$ (Eq.~(\ref{eqn:f})) for parameters of the wind 
$v_{\infty} = 20-50$\kms, $a = 1$\kms, and $\beta = 2.5$ 
(Sect.~2.3), and the equatorial rotation velocity of the giant 
$v_{\rm rot} = 6-9.5$\kms\ (Sect.~2.4). Selected models are 
summarized in Table~2. Models A--D correspond to 
$v_{\rm rot} \sim 0.9\times v_{\rm rot,max}$, while models 
E--H have $v_{\rm rot} \ga 0.95\times v_{\rm rot,max}$. 

Figure~\ref{fig:f} shows $f$, for $r < 6\,R_{\star}$, all 
$\theta$ and parameters of models A and H in Table~2. 
Compression to the equatorial plane is fairly 
significant: $f > 1$ for $\theta \ga 75{\degr}$ and 
$f \sim 5-10$ for $\theta = 90{\degr}$ at $r = 5-6\,R_{\star}$. 
This means that the wind from slowly rotating giants is 
compressed to its equatorial plane with the factor 
$f \sim 5-10$ with respect to the density distribution 
of the spherically symmetric wind that has the same 
overall $\dot M$. 

Figure~\ref{fig:den} shows examples of the mass density 
distribution in the compressed wind calculated according 
to Eq.~(\ref{eqn:nh}) for typical quantities of RGs in 
S-type symbiotic stars, $\dot M = 10^{-7}$\myr, 
$R_{\star} = 100$\ro, and parameters of models A and H 
as in Fig.~\ref{fig:f}. A compression of the wind is 
clearly seen. 

Figure \ref{fig:mm} shows the dependence of $f$ on the radial 
distance at the equatorial plane (left and middle panels) and 
on the rotation rate $v_{\rm rot}/v_{\rm rot,max}$ (right panel). 
For the fixed $v_{\infty}$ the density gradient in the 
direction perpendicular to the equatorial plane is larger 
for higher $v_{\rm rot}$. 
This implies an increase of $f$ when $v_{\rm rot}$ approaches 
its threshold value $v_{\rm rot,max}$ for forming a WCZ, and 
thus a more efficient wind mass transfer (the right panel of 
the figure). 

Our results suggest that for a typical separation between 
the binary components of 5--6\,$R_{\star}$ (2--3 AU), 
the wind compression can increase the local mass-loss 
rate around the orbital plane, and thus also the 
accretion rate onto the WD, by a factor of $\sim 5-10$ 
with respect to the spherically symmetric wind. 

\section{Discussion}
Because of the low mass transfer ratio in the wind accreting 
binaries (see Sect.~1), the only way to get the required 
high accretion rate, which maintains the high luminosity of 
WDs in symbiotic systems, is to focus the giant wind in 
the direction of their compact companions. 
Our findings that the rotating giant can compress its wind 
around the WD by a factor of $f \sim 5-10$ 
(Figs.~\ref{fig:f} and \ref{fig:mm}) suggests a relevant 
increase in the accretion rate, $\dot M_{\rm acc}$, onto 
the WD. It can be formally expressed as 
\begin{equation}
  \dot M_{\rm acc} \approx \eta \times f \times \dot M, 
\label{eqn:macc}
\end{equation}
where $\eta$ is the mass transfer ratio (a few times 0.01 as 
summarized in Sect.~1), $f$ is the mass-loss ratio 
(Eq.~(\ref{eqn:f}), Fig.~\ref{fig:mm}) 
and $\dot M$ is the mass-loss rate from giants in S-type 
symbiotic stars. 
Thus, $\dot M = {\rm a~few}~\times 10^{-7}$\myr\ 
yields 
$\dot M_{\rm acc} \sim 10^{-8} - 10^{-7}$\myr, which is 
sufficient to power the WD luminosity by a stable 
hydrogen-burning regime for the WD mass $< 1$\mo. 

The WRLOF represents a different type of the wind focusing 
(Sect.~1), which has recently been applied to the nearest 
D-type symbiotic binary, Mira AB (o~Ceti), to explain its 
unusual activity \citep[][]{moh+pod12}. It was shown that 
the WRLOF mode is applicable for slow dense winds from 
evolved AGB objects with the acceleration zone extending to 
their Roche lobes \citep[see Fig.~1 of][]{abate+13}. 
Measuring and modelling the circumstellar envelope around 
o~Ceti, the prototype of Mira variables, showed that the 
terminal expansion velocity of its wind is only 
4.6 -- 2.5\kms\ \citep[][]{b+k88,r+s01}. 
This suggests that the gravitational focusing by the WRLOF 
mode may also be in the effect in other D-type symbiotic 
stars, because they comprise a Mira-type variable as 
the cool component. 

On the other hand, the winds from normal giants in S-type 
symbiotic stars are typically one order of magnitude faster 
than those produced by Mira variables. 
Therefore, focusing of the wind in S-type symbiotic stars 
by the rotation of their normal giants probably represents 
a more suitable alternative to the WRLOF. 
However, this idea should be tested further by applying 
the WRLOF mode also to faster winds to better understand 
which type of the mass transfer mode is dominant in 
different wide-interacting binaries. 

\section{Conclusion}

We tested a possibility of the wind focusing towards the orbital 
plane of S-type symbiotic binaries by rotation of their red 
giants. In this way we attempted to explain a high wind mass 
transfer efficiency required to power the observed luminosity 
of their WD accretors. 

For the measured rotational velocities of the giant, 6--10\kms, 
and terminal velocities of its wind, 20--50\kms, the densities 
of the compressed wind around the WD, hence the accretion 
rate, can be a factor of $\sim 10$ higher than in the case of 
spherically symmetric wind (Figs.~\ref{fig:f} and \ref{fig:mm}, 
Eq.~(\ref{eqn:macc})). 
As a result, the WD can accrete at rates of $10^{-8}-10^{-7}$\myr\ 
(Sect.~4), which is sufficient to power its luminosity of a few 
times 1000\lo\ measured during quiescent phases of symbiotic 
stars (Sect.~1). 

Being aware of the simplicity of the WCD model for the purpose of 
our application, more sophisticated calculations should be 
done to prove (or disprove) our result. 
Using the WCD model we can suggest only the way to get enough 
wind from the RG to the vicinity of the WD in S-type symbiotic 
binaries. However, to find out the structure of the wind from 
the rotating giant in these systems and to understand 
the process of its accretion, 3-D hydrodynamic models 
(e.g., as cited in Sect.~1 and/or that was recently presented 
by \cite{hadcech12}), including the idea of the WCD model, 
should gone into in more detail. 
From this point of view, our approach to solving the problem of
a high wind mass transfer efficiency in S-type symbiotic binaries 
represents a direction for future theoretical modelling. 

\begin{acknowledgements}
The authors would like to thank the anonymous referee 
for useful comments that helped to improve the clarity 
of this paper. 
This research made use of NASA's Astrophysics Data System 
Service. 
This project was supported by the Slovak Academy of Sciences 
under grant VEGA No.~2/0002/13.

\end{acknowledgements}


\begin{thebibliography}{}
%
\bibitem[Abate et al. (2013)]{abate+13}
         Abate, C., Pols, O.R., Izzard, R.G., Mohamed, S.S.,
         de Mink, S. E., 2013. A\&A 552, A26.
%
\bibitem[Belczynski et al. (2000)]{belcz+00}
         Belczynski, K., Mikolajewska, J., Munari, U., Ivison, R. J.,
         Friedjung, M. 2000, Astron. Astrophys. Suppl. Ser., 146, 407
%
\bibitem[Bjorkman \& Cassinelli (1993)]{bjorkcass93}
         Bjorkman, J. E., Cassinelli, J.P. 1993, ApJ, 409, 429
%
\bibitem[Blind et al. (2011)]{blind+11}
         Blind, N., Boffin, H. M. J., Berger, J.-P., et al.
         2011, A\&A, 536, A55
%
\bibitem[Boffin et al. (2014a)]{boffin+14a}
         Boffin, H. M. J., Blind, N., Hillen, M., Berger, J.-P.,
         Jorissen, A., \& Le Bouquin, J.-B. 
         2014, Messenger 156  
%
\bibitem[Boffin et al. (2014b)]{boffin+14b}
         Boffin, H. M. J., Hillen, M., Berger, J. P. et al.
         2014, A\&A, 564, A1
%
\bibitem[Bondi \& Hoyle (1944)]{b+h44}
         Bondi, H., \& Hoyle, F. 1944, MNRAS, 114, 195
%
\bibitem[Bowers \& Knapp (1988)]{b+k88}
         Bowers, P. F., \& Knapp, G. R. 
         1988, ApJ, 332, 299
%
\bibitem[Carikov\'a \& Skopal (2014)]{zcsk14}
         Carikov\'a, Z., Skopal, A. 2014, A\&A, 570, A4
%
\bibitem[de Val-Borro et al. (2009)]{borro+09}
         de Val-Borro, M., Karovska, M., \& Sasselov, D.
         2009, ApJ, 700, 1148
%
\bibitem[Dumm et al. (2000)]{dumm+00}
         Dumm, T., Folini, D., Nussbaumer, H., et al. 
         2000, A\&A, 354, 1014
%
\bibitem[Hadrava \& \v{C}echura (2012)]{hadcech12}
         Hadrava, P., \& \v{C}echura, J. 2012, A\&A, 542, A42
%
\bibitem[Ignace et al. (1996)]{igncassbjork96}
         Ignace, R., Cassinelli, J. P., Bjorkman, J. E. 
         1996, ApJ, 459, 671
%
%
\bibitem[Kenyon \& Gallagher (1983)]{ken+gall83}
         Kenyon, S. J., \& Gallagher, J. S.
         1983, AJ, 88, 666
%
\bibitem[Lamers \& Cassinelli (1999)]{lamcass99}
         Lamers, H. J. G. L. M., Cassinelli, J. P. 
         1999, Introduction to stellar winds, 
         Cambridge University Press
%
\bibitem[Livio \& Warner (1984)]{livwar84}
         Livio, M., Warner, B. 
         1984, The Observatory, 104, 152
%
\bibitem[Matthews \& Karovska (2006)]{matt+kar06}
         Matthews, L. D., \& Karovska, M. 2006, ApJ, 637, L49
%
\bibitem[Miko\l ajewska et al. (2002)]{mio02}
         Miko\l ajewska, J., Ivison, R. J., \& Omont, A. 2002,
         Adv. Space Res., 30, 2045
%
\bibitem[Mohamed \& Podsiadlowski (2007)]{moh+pod07}
         Mohamed, S., \& Podsiadlowski, Ph. 
         2007, In: $15^{th}$ European Workshop on White Dwarfs, 
         eds. R. Napiwotzki and M. R. Burleigh, ASP Conf. Ser. 
         Vol. 372, p. 397
%
\bibitem[Mohamed \& Podsiadlowski (2012)]{moh+pod12}
         Mohamed, S., \& Podsiadlowski, Ph. 
         2012, Baltic Astronomy, 21, 88
%
\bibitem[M\"urset et al. (1991)]{m+91}
         M\"urset, U., Nussbaumer, H., Schmid, H. M., \& Vogel, M.
         1991, A\&A, 248, 458
%
\bibitem[M\"urset \& Schmid (1999)]{ms99}  
         M\"urset, U., \& Schmid, H. M. 1999, A\&AS, 137, 473
%
\bibitem[Nagae et al. (2004)]{nagae+04}
         Nagae, T., Oka, K., Matsuda, T., Fujiwara, H., 
         Hachisu, I., \& Boffin, H. M. J. 
         2004, A\&A, 419, 335
%
\bibitem[Parimucha et al. (2002)]{parimucha+02}
         Parimucha, \v{S}., Chochol, D., Pribulla, T., 
         Buson, L. M., \& Vittone, A. A.
         2002, A\&A, 391, 999
%
\bibitem[Ryde \& Sch\"{o}ier (2001)]{r+s01}
         Ryde, N., \& Sch\"{o}ier, F. L. 
         2001, ApJ, 547, 384
%
\bibitem[Schild et al. (1996)]{schild+96}
         Schild, H., M\"urset, U., \& Schmutz, W. 
         1996, A\&A, 306, 477
%
\bibitem[Schr\"oder (1985)]{schroder}
         Schr\"oder, K.-P. 1985, A\&A, 147, 103
%
\bibitem[Shen \& Bildsten (2007)]{shen+07}
         Shen, K. J., \& Bildsten, L. 2007, ApJ, 660, 1444
%
\bibitem[Seaquist et al. (1984)]{stb}
         Seaquist, E. R., Taylor, A. R., \& Button, S. 1984,
         ApJ, 284, 202
%
\bibitem[Skopal (2005)]{sk05}
         Skopal, A. 2005, A\&A, 440, 995
%
\bibitem[Soker (2002)]{soker02}
         Soker, N. 2002, MNRAS, 337, 1038
%
\bibitem[Theuns et al. (1996)]{tbj96}
         Theuns, T., Boffin, H. M. J., \& Jorissen, A.
         1996, MNRAS, 280, 1264
%
\bibitem[Tutukov \& Yungelson (1976)]{tutukov+76}
         Tutukov, A. V., \& Yungelson, L. R. 1976, 
         Astrophysics, 12, 342
%
\bibitem[Walder (1997)]{walder97}
         Walder, R. 1997, in Accretion phenomena and related 
         outflows, eds. D. T. Wichramasinghe, L. Ferrario, \& 
         G. V. Bicknell, IAU Coll., 163, ASP Conf. Ser., 121, 
         San Francisco: ASP, 822
%
\bibitem[Walder et al. (2008)]{walder+08}
         Walder, R., Folini, D., \& Shore, S. N. 
         2008, A\&A, 484, L9
%
\bibitem[Webster \& Allen (1975)]{wa75}  
         Webster, B. L. \& Allen, D. A. 1975, MNRAS, 171, 171
%
\bibitem[Zamanov et al. (2007)]{zam+07}
         Zamanov, R. K., Bode, M. F., Melo, C. H. F. et al. 
         2007, MNRAS, 380, 1053
%
\bibitem[Zamanov et al. (2008)]{zam+08}
         Zamanov, R. K., Bode, M. F., Melo, C. H. F. et al. 
         2008, MNRAS, 390, 377 
%
\bibitem[Zamanov \& Stoyanov (2012)]{zamsto12}
         Zamanov, R. K., Stoyanov, K. A. 
         2012, Bulgarian Astron. J., 18(3), 41
%
\end{thebibliography}
\end{document}